\documentclass[12pt,a4paper]{article}

\usepackage[fleqn]{amsmath}
\usepackage{alltt}
\usepackage{amssymb}
\newtheorem{definition}{Definition}
\newtheorem{theorem}{Theorem}
\newtheorem{lemma}[theorem]{Lemma}

\def\Tr{{\rm Tr}}
\newcommand{\tfr}[1]{{\widehat{#1}}}
\begin{document}

\title{On the Fourier Spectra of the Infinite Families of Quadratic APN Functions}

\author{Carl Bracken$^1$, Zhengbang Zha$^2$\\
$^1$Department of Mathematics, National University of Ireland\\
Maynooth, Co. Kildare\\
$^2$College of Mathematics and Econometrics, Hunan University\\
Changsha  410082, China}

\maketitle

\bigskip
\begin{abstract}
\noindent
It is well known that a quadratic function defined on a finite field of odd degree is
almost bent (AB) if and only if it is almost perfect nonlinear (APN).
For the even degree case there is no apparent relationship between the values in the Fourier spectrum
of a function and the APN property.
In this article we compute the Fourier spectrum of the quadranomial family of APN functions from \cite{BB1}.
With this result, all known infinite families of APN functions now have their Fourier spectra and hence their nonlinearities computed.


\end{abstract}



\bigskip


\section{Introduction}
Highly nonlinear functions on finite fields are interesting from the point of view of cryptography as they provide optimum resistance to linear and differential attacks. A function that has the APN (resp. AB) property, as defined below, has optimal resistance to a differential (resp. linear) attack.
For more on relations between linear and differential cryptanalysis, see \cite{CV}.

Highly nonlinear functions are also of interest from the point of view of coding theory.
The weight distribution of a certain error-correcting code is equivalent to the
Fourier spectrum (including multiplicities) of $f$.
The code having three particular weights is equivalent to the AB property, when $n$ is odd.
The minimum distance of the dual code being 5 is equivalent to the APN property holding for $f$.

For the rest of the paper, let $L = GF(2^n)$
and let $L^*$ denote the set of non-zero elements of $L$.
Let $\Tr: L \rightarrow GF(2)$ denote the trace map from $L$ to $GF(2)$.

\begin{definition}
A function $f: L \rightarrow L$ is said to be {\emph{\bf{ almost perfect nonlinear (APN)}}}
if for any $a\in L^*, b \in L$,  we have
$$ |\{x \in L : f(x+a)-f(x) = b \}| \leq 2. $$
\end{definition}

\begin{definition}
Given a function $f:L \rightarrow L$, the {\emph {\bf{Fourier transform}}} of $f$ is the function
$\tfr f: L \times L^* \rightarrow \mathbb Z$ given by
$$\tfr f(a,b) = \sum_{x \in L}(-1)^{\Tr(ax+bf(x))}.$$
\label{WT}
\end{definition}

The {\emph{Fourier spectrum}} of $f$ is the set of integers
$$\Lambda_f= \{\tfr f(a,b) : a, b \in L, b \neq 0 \}.$$

The nonlinearity of a function $f$ on a field $L=GF(2^n)$ is defined as

$$NL(f):= 2^{n-1} - \frac{1}{2}\max_{x \in \Lambda_f} \ |x| .$$

The nonlinearity of a function measures its distance
to the set of all affine maps on $L$.
We thus call a function {\emph{maximally nonlinear}} if its nonlinearity is as large as possible.
If $n$ is odd, its nonlinearity is upper-bounded by $2^{n-1}-2^{\frac{n-1}{2}}$, while for $n$ even a conjectured upper bound is  $2^{n-1}-2^{\frac{n}{2}-1}$.
For odd $n$, we say that a function $f : L\longrightarrow L$ is \emph{almost bent} (AB)
when its Fourier spectrum is $\{0, \pm 2^{\frac{n+1}{2}}\} $,
in which case it is clear from the upper bound that $f$ is maximally nonlinear.
We have the following connection (for odd $n$)
between the AB and APN property: every AB
function on $L$ is also APN \cite{CV}, and, conversely,
if $f$ is quadratic and APN, then $f$ is AB \cite{CCZ}. 
In particular, quadratic APN functions have optimal resistance to both linear and differential attacks.
On the other hand, there appears to be no relation between the nonlinearity $NL(f)$
and the APN property of a function $f$ when $n$ is even.
The reader is referred to \cite{C} for a comprehensive survey on APN and AB functions.

\section{New Families of Quadratic APN functions}

Recently, the first non-monomial families of APN functions have been discovered.
Below we list the new families of non monomial functions known at the time of writing.

\bigskip

\begin{enumerate}
\item
$$f(x)= x^{2^s+1}+ \alpha x^{2^{ik}+2^{mk+s}},$$
where $n =3k$, $(k,3)=(s,3k)=1$, $k \geq 3$, $i \equiv sk \mod 3$, $m \equiv -i \mod 3$, $\alpha = t^{2^k-1}$ and $t$ is primitive (see
 Budaghyan,  Carlet,   Felke,  Leander \cite{BCFL}).

\bigskip

\item

$$f(x)= x^{2^s+1}+ \alpha x^{2^{ik}+2^{mk+s}},$$ where
$n =4k$, $(k,2)=(s,2k)=1$, $k \geq 3$, $i \equiv sk \mod 4$, $m= 4-i$, $\alpha = t^{2^k-1}$ and $t$ is primitive (see Budaghyan,  Carlet,  Leander \cite{BCL2}).
This family generalizes an example found for $n=12$ by
 Edel, Kyureghyan, Pott \cite{EKP}.

\bigskip

\item
$$f(x) = \alpha x^{2^{s}+1}+{\alpha}^{2^k}x^{2^{k+s}+2^k}+\beta
x^{2^k+1}+\sum_{i=1}^{k-1} {\gamma}_i x^{2^{k+i}+2^i},$$
where $n=2k$,
$\alpha$ and $\beta$ are primitive elements of $GF(2^{n})$, and
${\gamma}_i \in GF(2^k)$ for each $i$, and $(k,s)=1$, $k$ is odd,
$s$ is odd (see  Bracken,  Byrne,  Markin,  McGuire \cite{BBMMcG}).

\bigskip

\item
$$f(x) = x^3 + Tr(x^9),$$
over $GF(2^n),$ any $n$ (see Budaghyan,  Carlet,  Leander  \cite{B+}).

\bigskip

\item
$$f(x)={\alpha}^{2^k}x^{2^{-k}+2^{k+s}}+{\alpha}x^{2^{s}+1}+vx^{2^{-k}+1}+w{\alpha}^{{2^k}+1}x^{2^{k+s}+2^s}$$
where $n=3k$,
$\alpha$ is primitive in $GF(2^n)$, $v,w \in GF(2^k)$ and $vw \neq 1$, $(s,3k)=1,\  (3,k)=1$ and 3 divides $k+s$
(see Bracken,  Byrne,  Markin,  McGuire \cite{BB1}).
\end{enumerate}

In  \cite{BBMMc1} the Fourier spectra of families (1) and (2) are computed.
The determination of the Fourier spectra of families (3) and (4)
has been given in \cite{BBMMcG2} and \cite{BBMMcG3}, respectively.
In this paper we calculate the Fourier spectra of family (5).
We will show here that the Fourier spectra of this family of functions
are 5-valued $\{0, \pm 2^{\frac{n}{2}}, \pm 2^{\frac{n+2}{2}} \}$ for fields of even degree and 3-valued  $\{0, \pm 2^{\frac{n+1}{2}}\} $ for fields of odd degree.
In this sense they resemble the Gold functions $x^{2^d+1}$, $(d,n)=1$, as indeed do all five APN functions listed above.
For fields of odd degree, our result provides another proof of the APN property.
This does not hold for fields of even degree;
as we stated earlier,
there appears to be no relation between the Fourier spectrum and the APN property
for fields of even degree.
Thus, the fact that $f$ has a $5$-valued Fourier
spectrum for fields of even degree does not follow from the fact that $f$ is a quadratic APN function.
Indeed, there is one example known (due to Dillon \cite{Dillon})
of a quadratic APN function on a field of even degree whose
Fourier spectrum is more than 5-valued;
if $u$ is primitive in $GF(2^6)$ then
$$g(x)=x^3+u^{11}x^5+u^{13}x^9+x^{17}+u^{11}x^{33}+x^{48}$$
is a quadratic APN function on $GF(2^6)$ whose Fourier transform takes seven distinct values.

\section{The Fourier Spectrum of Family (5)}
We shall make use of the following lemma, a proof of which can be found in \cite{BBMMc1}.
\begin{lemma}
Let $s$ be an integer satisfying $(s,n)=1$ and let
$f(x) =\displaystyle{ \sum_{i=0}^d r_i x^{2^{si}}}$ be a polynomial in $ L[x]$. Then $f(x)$ has at most $2^d$ zeroes in $L$.
\label{rsol}
\end{lemma}

\begin{theorem}
Let $f(x)={\alpha}^{2^k}x^{2^{-k}+2^{k+s}}+{\alpha}x^{2^{s}+1}+vx^{2^{-k}+1}+w{\alpha}^{{2^k}+1}x^{2^{k+s}+2^s}$, where $n=3k$,
$\alpha$ is primitive in $GF(2^n)$, $v,w \in GF(2^k)$ and $vw \neq 1$, $(s,3k)=1,\  (3,k)=1$ and 3 divides $k+s$
The Fourier spectrum of $f(x)$ is $\{0, \pm 2^{\frac{n+1}{2}}\} $ when $n$ is odd and $\{0, \pm 2^{\frac{n}{2}}, \pm 2^{\frac{n+2}{2}} \}$
when $n$ is even.
\end{theorem}

\noindent{\bf Proof:}
The Fourier spectrum of $f$ is given by

$$\tfr f(a,b) = \sum_{x \in L}(-1)^{Tr(ax+ b f(x))}. $$
Squaring gives
\begin{eqnarray*}
\tfr f(a,b)^2  & = & \sum_{x \in L} \sum_{y \in L}(-1)^{\Tr(ax+bf(x)+ay+bf(y))}\\
            & = & \sum_{x \in L} \sum_{u \in L}(-1)^{\Tr(ax+bf(x)+a(x+u)+bf(x+u))},
\end{eqnarray*}
from the substitution $y=x+u$.

This becomes
$$\tfr f(a,b)^2 = \sum_u{(-1)^{Tr(au+bf(u))}}\sum_x{(-1)^{Tr(xL_b(u))}},$$
where $$L_b(u): ={\alpha}bu^{2^s}+{\alpha}^{2^{-s}}b^{2^{-s}}u^{2^{-s}}
+{\alpha}^{2^{-k}}b^{2^k}u^{2^{-k+s}} + {\alpha}^{2^{-s}}b^{2^{-k-s}}u^{2^{k-s}}$$
$$ + vb^{2^k}u^{2^k}+ vbu^{2^{-k}} + w^{2^{-s}}b^{2^{-k-s}}{\alpha}^{2^{-s}+2^{-k-s}}u^{2^{-k}}
+ w^{2^{-s}}b^{2^{-s}} {\alpha}^{2^{k-s}+2^{-s}}u^{2^k} .$$

Using the fact that $ \sum_x(-1)^{Tr(cx)}$ is $0$ when $c \neq 0$ and $2^n$ otherwise, we obtain

$$\tfr f(a,b)^2 = 2^n \sum_{u \in K} (-1)^{Tr(au+bf(u)},$$

where $K$ denotes the kernel of $L_b(u)$. If the size of the kernel is at most 4,
then clearly

$$0 \leq \sum_{u \in K} (-1)^{Tr(au+b f(u))} \leq 4.$$

Since  $\tfr f(a,b)$ is an integer, this sum can only be 0, 2, or 4 if $n$ is even, and 1 or 3 if $n$ is odd.
 The set of permissible values of $\tfr f(a,b)$  is then

$\{0, 2^{\frac{n+1}{2}},-2^{\frac{n+1}{2}} \}$ when $n$ is odd and $\{ 0, 2^{n/2}, -2^{n/2}, 2^{\frac{n+2}{2}},-2^{\frac{n+2}{2}} \}$ when $n$ is even.

$$\tfr f(a,b) \in
\begin{cases}
\{0, \pm 2^{\frac{n+1}{2}} \} & 2 \nmid n \\
\{ 0, \pm 2^{\frac{n}{2}}, \pm 2^{\frac{n+2}{2}} \} & 2 \mid n.
\end{cases}
$$

We must now demonstrate that $|K| \leq 4$, which is sufficient to complete the proof.

Now suppose that $L_b(u)=0$. This gives
$$b^{2^{-k}}L_b(u)=b^{2^{-k}-2^{k-s}}{\alpha}^{2^{-s}}(b^{{2^{-s}}+{2^{k-s}}}u^{2^{-s}} + b^{{2^{k-s}}+{2^{-k-s}}}u^{2^{k-s}}$$
$$ + w^{2^{-s}}b^{2^{-s}+{2^{k-s}}} {\alpha}^{2^{k-s}}u^{2^k}
+  w^{2^{-s}}b^{2^{k-s}+{2^{-k-s}}} {\alpha}^{2^{-k-s}}u^{2^{-k}}) $$
$$\ \ \ \  +  {\alpha}^{2^{-k}}b^{{2^{k}}+{2^{-k}}}u^{2^{-k+s}} + {\alpha}b^{{2^{-k}}+1}u^{2^s}
+  vb^{{2^{k}}+{2^{-k}}}u^{2^{k}} + vb^{{2^{-k}}+1}u^{2^{-k}} =0.\ \ \ \  \ \ (1)$$
Next we let $\theta = {\alpha}^{2^{-s}}b^{2^{-k}-2^{k-s}}$, \ $t(u)
= b^{{2^{-s}}+{2^{k-s}}}(u^{2^{-s}} +
w^{2^{-s}}{\alpha}^{2^{k-s}}u^{2^k} )$ and $r(u) =
b^{{2^{k}}+{2^{-k}}}(vu^{2^{k}} + {\alpha}^{2^{-k}}u^{2^{-k+s}}).$
Equation (1) now becomes
$$\ \ \ \ \ \ \ \ \ \ \ \ \  \ \ \ \ \ \ b^{2^{-k}}L_b(u)=r(u)+r(u)^{2^k}+\theta(t(u)+t(u)^{2^k})=0 \ \ \ \ \ \ \ \ \ \ \ \ \ \ \ \ \ \ \ \ (2)$$
For convenience we will write $r(u)$ and $t(u)$ as $r$ and $t$. We have,
$$t^{2^{k+s}}=b^{2^{k}+2^{-k}}(u^{2^k}+w{\alpha}^{2^{-k}}u^{2^{-k+s}}).$$
This implies
$$\ \ \ \ \ \ \ \ \ \ \ \ \ \ \ \ \ \ \ \ \ t^{2^{k+s}}+wr=b^{2^{k}+2^{-k}}(1+vw)u^{2^k}.\ \ \ \ \  \ \ \ \ \ \ \ \ \ \ \ \ \ \ \ \ \ \ \ (3)$$
We also get
$$\ \ \ \ \ \ \ \ \ \ \ \ \ \ \ \ \ vt^{2^{k+s}}+r=b^{2^{k}+2^{-k}}(1+vw){\alpha}^{2^{-k}}u^{2^{-k+s}}. \ \ \ \ \ \ \ \ \ \ \ \ \ \ \ \ \ \ \ (4)$$
Equation (3) implies
$$u=b^{-1-2^k}(1+vw)^{-1}(t^{2^s}+wr^{2^{-k}}),$$
while Equation (4) gives
$$u=b^{-2^{-k-s}-2^{-s}}(1+vw)^{-2^{-s}}{\alpha}^{-2^{-s}}(v^{2^{-s}}t^{2^{-k}}+r^{2^{k-s}}).$$
Combining these two expressions for $u$ yields
$$\theta z(t^{2^s}+wr^{2^{-k}})=v^{2^{-s}}t^{2^{-k}}+r^{2^{k-s}},$$
where $z=(1+vw)^{2^{-s}-1}(b^{2^{-k}+2^{k}+1})^{2^{-s}-1}$. Note $z \in GF(2^k)$.
We rearrange and multiply by $\theta + \theta^{2^{-k}}$ to obtain
$$\ \ \ \ \ \ \ \ \ (\theta + \theta^{2^{-k}})(wz\theta^{2^k} r+v^{2^{-s}}t)=(\theta + \theta^{2^{-k}})(\theta^{2^k} zt^{2^{k+s}}+r^{2^{-k-s}}). \ \ \ \ \ \ \ \ \ (5)$$
We claim that $\theta + \theta^{2^{-k}}$ is not zero. If $\theta = \theta^{2^{-k}}$ then
$\alpha^{2^k-1}=b^{(2^{k+s}-1)(2^{k}-2^{-k})}$. As $k+s$ is divisible by 3, $2^{k+s}-1$ is divisible by seven. This implies $\alpha$ is a seventh power contradicting its primitive status and the claim is proven.
From Equation (2) we have
$$r+r^{2^k}=\theta(t+t^{2^k})$$
From this equation and using the fact that relative trace mapping from $GF(2^{3k})$ to  $GF(2^{k})$ (denoted by $Tr_k$) is zero for any field element of the form $\delta+\delta^{2^k}$, we derive the following
$$Tr_k(\theta(t+t^{2^k})) = Tr_k(\theta^{-1}(r+r^{2^k}))=0.$$
As $Tr_k(cg)=c Tr_k(g)$ for $c \in GF(2^k)$ and $\theta^{2^{-k}+2^k+1} \in GF(2^k)$ we can say
$$\ \ \ \ \ \ \ \ \ \ \ \ \ \ \ \ \ \ \  Tr_k((\theta + \theta^{2^{-k}})t) = Tr_k(\theta^{2^k}(\theta + \theta^{2^{-k}})r)=0.\ \ \ \ \ \ \ \ \ \ \ \ \ \ \ \ \ \ \ (6)$$
Therefore the left hand side of Equation (5) has relative trace of
zero, which implies the right hand side of Equation (5) has relative
trace of zero also. That is,
$$Tr_k((\theta + \theta^{2^{-k}})(\theta^{2^k} zt^{2^{k+s}}+r^{2^{-k-s}}))=0.$$
We write this as
$$z((\theta + \theta^{2^{-k}})\theta^{2^k} t^{2^{k+s}}+(\theta + \theta^{2^{-k}})^{2^k}\theta^{2^{-k}} t^{2^{-k+s}}+(\theta + \theta^{2^{-k}})^{2^{-k}}\theta t^{2^{s}})$$
$$\ \ \ \ \ \ \ \ =(\theta + \theta^{2^{-k}})r^{2^{-k-s}}+(\theta + \theta^{2^{-k}})^{2^{k}}r^{2^{-s}}+(\theta + \theta^{2^{-k}})^{2^{-k}}r^{2^{k-s}}.\ \ \ \ \ \ \ \ \ (7)$$
From Equation (6) we obtain
$$t^{2^{-k+s}}=(\theta + \theta^{2^{-k}})^{2^{s}-2^{-k+s}}t^{2^s}+(\theta + \theta^{2^{-k}})^{2^{k+s}-2^{-k+s}}t^{2^{k+s}},$$
$$r^{2^{-k-s}}=\theta^{2^{k-s}-2^{-s}}(\theta + \theta^{2^{-k}})^{2^{-s}-2^{-k-s}}r^{2^{-s}}+\theta^{2^{-k-s}-2^{-s}}(\theta + \theta^{2^{-k}})^{2^{k-s}-2^{-k-s}}r^{2^{k-s}}.$$
Substituting these expressions for $t^{2^{-k+s}}$ and $r^{2^{-k-s}}$ into Equation (7) we get
$$z(((\theta + \theta^{2^{-k}})\theta^{2^{k}}+(\theta + \theta^{2^{-k}})^{2^{k}}\theta^{2^{-k}}(\theta + \theta^{2^{-k}})^{2^{k+s}-2^{-k+s}})t^{2^{k+s}} +$$
$$ ((\theta + \theta^{2^{-k}})^{2^{-k}}\theta+(\theta + \theta^{2^{-k}})^{2^{k}}\theta^{2^{-k}}(\theta + \theta^{2^{-k}})^{2^{s}-2^{-k+s}})t^{2^s})$$
$$= ((\theta + \theta^{2^{-k}})^{2^k}+(\theta + \theta^{2^{-k}})\theta^{2^{k-s}-2^{-s}}(\theta + \theta^{2^{-k}})^{2^{-s}-2^{-k-s}})r^{2^{-s}} +$$
$$((\theta + \theta^{2^{-k}})^{2^{-k}}+(\theta + \theta^{2^{-k}})\theta^{2^{-k-s}-2^{-s}}(\theta + \theta^{2^{-k}})^{2^{k-s}-2^{-k-s}})r^{2^{k-s}}.$$
We multiply across by $(\theta +
\theta^{2^{-k}})^{2^{-k-s}+2^{-k+s}}\theta^{2^{-s}}$ and obtain
$$z(\theta + \theta^{2^{-k}})^{2^{-k-s}}\theta^{2^{-s}}(((\theta + \theta^{2^{-k}})^{2^{-k+s}+1}\theta^{2^{k}}+(\theta + \theta^{2^{-k}})^{2^{k+s}+2^{k}}\theta^{2^{-k}})t^{2^{k+s}}$$
$$+((\theta + \theta^{2^{-k}})^{2^{-k+s}+2^{-k}}\theta+(\theta + \theta^{2^{-k}})^{2^{s}+2^{k}}\theta^{2^{-k}})t^{2^{s}})=$$
$$(\theta + \theta^{2^{-k}})^{2^{-k+s}}(((\theta + \theta^{2^{-k}})^{2^{-k-s}+2^{k}}\theta^{2^{-s}}+(\theta + \theta^{2^{-k}})^{2^{-s}+1}\theta^{2^{k-s}})r^{2^{-s}}$$
$$+((\theta + \theta^{2^{-k}})^{2^{-k-s}+2^{-k}}\theta^{2^{-s}}+(\theta + \theta^{2^{-k}})^{2^{k-s}+1}\theta^{2^{-k-s}})r^{2^{k-s}}).$$
Letting $P(\theta)=(\theta + \theta^{2^{-k}})^{2^{-k+s}+1}\theta^{2^{k}}+(\theta + \theta^{2^{-k}})^{2^{k+s}+2^{k}}\theta^{2^{-k}}$, the above equation becomes
$$z(\theta + \theta^{2^{-k}})^{2^{-k-s}}\theta^{2^{-s}}(P(\theta)t^{2^{k+s}}+P(\theta)^{2^{-k}}t^{2^{s}})$$
$$\ \ \ \ \ \ \ \ \ \ \ \ \ \ \ \ \ \ \ \ \ \ =(\theta + \theta^{2^{-k}})^{2^{-k+s}}(P(\theta)^{2^{-k-s}}r^{2^{-s}}+P(\theta)^{2^{k-s}}r^{2^{k-s}}).\ \ \ \ \ \ \ \ \ \ \ \ \ \ \ \ \ \ \ \ (8)$$
We claim that $P(\theta)$ is a non zero element of $GF(2^k)$.
Setting $P(\theta)$ equal to zero yields
$$(\theta + \theta^{2^{-k}})^{2^{-k+s}+1}\theta^{2^{k}}=(\theta + \theta^{2^{-k}})^{2^{k+s}+2^{k}}\theta^{2^{-k}}.$$
This implies
$$\theta^{2^{k}-2^{-k}}=(\theta + \theta^{2^{-k}})^{(2^{k+s}-1)(1-2^{k})}.$$
Therefore $\theta^{2^{k}-2^{-k}}$ is a seventh power. But
$\theta^{2^{k}-2^{-k}}=(\alpha b^{2^k(2^{k+s}-1)})^{2^{k}-2^{-k}}$, which would require $\alpha$ to be a seventh power also, which its not. Hence $P(\theta) \neq 0$.

To see that $P(\theta) \in GF(2^k)$, we multiply the expression out and refactor as follows.
$$P(\theta)= (\theta + \theta^{2^{-k}})^{2^{-k+s}}(\theta + \theta^{2^{-k}})\theta^{2^{k}}+
(\theta + \theta^{2^{-k}})^{2^{k+s}}(\theta + \theta^{2^{-k}})^{2^{k}}\theta^{2^{-k}}.$$
This implies
$$P(\theta)= (\theta^{2^{-k+s}} + \theta^{2^{k+s}})(\theta^{2^{k}+1} + \theta^{2^{-k}+2^{k}}) + (\theta^{2^{s}} + \theta^{2^{k+s}})(\theta^{2^{-k}+1} + \theta^{2^{-k}+2^{k}}),$$
which becomes
$$P(\theta)= \theta^{2^{s}}(\theta^{2^{-k}+1} + \theta^{2^{-k}+2^{k}}) +
\theta^{2^{k+s}}(\theta^{2^{k}+1} + \theta^{2^{-k}+1}) +
\theta^{2^{-k+s}}(\theta^{2^{k}+2^{-k}} + \theta^{2^{k}+1}). $$
We can write this as
$$P(\theta)=Tr_k(\theta^{2^{s}}(\theta^{2^{-k}+1} + \theta^{2^{-k}+2^{k}})),$$
hence $P(\theta) \in GF(2^k)$ and the claim is proven.

Now Equation (8) becomes
$$z(\theta + \theta^{2^{-k}})^{2^{-k-s}-2^{-k+s}}P(\theta)^{1-2^{-s}}(t+t^{2^{k}})^{2^{s}}
 = \theta^{-2^{-s}}(r+r^{2^k})^{2^{-s}}.$$
Using Equation (2) and raising by $2^s$ we obtain an equation in $(t+t^{2^k})$,
$$z^{2^s}(\theta + \theta^{2^{-k}})^{2^{-k}-2^{-k+2s}}P(\theta)^{2^{s}-1}(t+t^{2^{k}})^{2^{2s}}
 +(t+t^{2^k})=0,$$
 which by Lemma 1 can have no more than two solutions for $(t+t^{2^k})$ when $n$ is odd and no more than four solutions for $(t+t^{2^k})$ when $n$ is even. This restriction on $(t+t^{2^k})$
is crucial and will be used to complete the proof, but first we consider the following two expressions which come from Equations (3) and (4) respectively,
$$Bu=t^{2^s}+wr^{2^{-k}},$$
$$\alpha^{2^{-s}}B^{2^{-k-s}}u=v^{2^{-s}}t^{2^{-k}}+r^{2^{k-s}},$$
where $B=(1+vw)b^{2^{k}+1}.$
From these we obtain
$$Bu+B^{2^k}u^{2^k}=(t+t^{2^k})^{2^s}+w(r+r^{2^k})^{2^{-k}},$$
$$\alpha^{2^{-s}}B^{2^{-k-s}}u+\alpha^{2^{k-s}}B^{2^{-s}}u^{2^k}=v^{2^{-s}}(t+t^{2^k})^{2^{-k}}+(r+r^{2^k})^{2^{k-s}}.$$
Next we eliminate the $u^{2^k}$ term in these equations to give the following
$$(\alpha^{2^{k-s}}B^{2^{-s}+1}+\alpha^{2^{-s}}B^{2^{-k-s}+2^k})u
= B^{2^k}(v^{2^{-s}}(t+t^{2^k})^{2^{-k}}+(r+r^{2^k})^{2^{k-s}})$$
$$\ \\ \ \ \ \ \ \ \ \ \ \ \ \ \ \ \ \ \ \ \ + \alpha^{2^{k-s}}B^{2^{-s}}
((t+t^{2^k})^{2^s}+w(r+r^{2^k})^{2^{-k}}).\ \ \ \ \ \ \ \ \ \ \ \ \ \ \ \ \ \ \ \ \ \ \ \ \ \ (9)$$
We let $D=\alpha^{2^{k-s}}B^{2^{-s}+1}+\alpha^{2^{-s}}B^{2^{-k-s}+2^k}$ and note that $D$ is not zero as $D=0$ implies  $\alpha^{2^{k-s}-2^{-s}}=B^{2^{-k-s}+2^k-2^{-s}-1}=
B^{(2^{k+s}-1)(2^{-s}-2^{-k-s})}$, which again contradicts the fact that $\alpha$ is primitive.
Therefore we may write Equation (9) as
$$u=D^{-1}(B^{2^k}(v^{2^{-s}}(t+t^{2^k})^{2^{-k}}+(r+r^{2^k})^{2^{k-s}}) $$
$$+ \alpha^{2^{k-s}}B^{2^{-s}}
((t+t^{2^k})^{2^s}+w(r+r^{2^k})^{2^{-k}})).$$
We now use Equation (2) to substitute the $r+r^{2^k}$ terms for $\theta(t+t^{k})$ and we obtain
$$u=D^{-1}(B^{2^k}(v^{2^{-s}}(t+t^{2^k})^{2^{-k}}+(\theta(t+t^{k}))^{2^{k-s}})$$
$$ + \alpha^{2^{k-s}}B^{2^{-s}}
((t+t^{2^k})^{2^s}+w(\theta(t+t^{k}))^{2^{-k}})).$$
Recall $t+t^{2^k}$ can only take two values when $n$ is odd and four when $n$ is even, hence the above equation shows that $u$ must have at least the same restrictions and the proof is complete.

\end{document}